\documentclass[twocolumn,preprintnumbers,amsmath,amssymb]{revtex4-2}
\usepackage{graphicx}
\usepackage{dcolumn}
\usepackage{bm}
\usepackage{color}
\usepackage{braket}
\usepackage{float}
\usepackage{adjustbox}
\usepackage[caption=false]{subfig}
\usepackage{nicefrac}

\usepackage{dcolumn}
\newcommand{\kbar}{\mathchar'26\mkern-9mu k}
\newcommand{\Ez}{\langle E_z \rangle}

\usepackage{comment}
\usepackage{hyperref}
\hypersetup{
	colorlinks   = true, 
	urlcolor     = blue, 
	linkcolor    = blue, 
	citecolor   = blue 
}

\usepackage{lineno}

\begin{document}

\title{Observation of Many-body Dynamical Delocalization in a Kicked 1D Ultracold Gas}

\author{Jun Hui See Toh,$^{1}$ Katherine C. McCormick,$^{1}$ Xinxin Tang,$^1$ Ying Su,$^{2}$ Xi-Wang Luo,$^{2}$ Chuanwei Zhang,$^{2}$ and Subhadeep Gupta$^{1}$\\
$^{1}${\em Department of Physics, University of Washington, Seattle, Washington 98195, USA}\\
$^{2}${\em Department of Physics, The University of Texas at Dallas, Richardson, Texas 75080-3021, USA}}
\affiliation{}



\maketitle

{\bf 
Contrary to a driven classical system that exhibits chaos phenomena and diffusive energy growth, a driven quantum system can exhibit dynamical localization that features energy saturation. However, the evolution of the dynamically localized state in the presence of many-body interactions has long remained an open question. Here we experimentally study an interacting 1D ultracold gas periodically kicked by a pulsed optical lattice, and observe the interaction-driven emergence of dynamical delocalization and many-body quantum chaos. The observed dynamics feature a sub-diffusive energy growth manifest over a broad parameter range of interaction and kick strengths, and shed light on an area where theoretical approaches are extremely challenging.}

The classical kicked rotor is a textbook paradigm to explore chaos phenomena, displaying a diffusively growing kinetic energy proportional to time or kick number above a critical kick strength \cite{lich92,chir79}. Dynamical localization in the quantum kicked rotor (QKR) \cite{casa79,lema09} arises from quantum interference, and can be explained by mapping \cite{fish82,grem84} the Floquet dynamics of the QKR to a disordered Anderson model \cite{ande58} in the momentum space lattice. In the past three decades, single particle QKRs have been experimentally studied extensively with cold neutral atoms and dynamical localization has been observed \cite{moor94,moor95,amma98,darc01,wimb03,duff04,ulla12,gadw13}. 

Understanding the role of many-body interactions in a disordered quantum system has been a long-standing challenge since the discovery of Anderson localization \cite{ande58}. In recent years, many-body localization in disordered lattices in position space has been extensively studied both experimentally and theoretically by incorporating methods developed in quantum information science \cite{schr15,luki19,aban19}. Despite the equivalence of dynamical localization \cite{fish82,grem84} to Anderson localization \cite{ande58} for a single particle, the infinite long-range interaction in the momentum-space lattice is fundamentally different from the short-range interaction in position-space Anderson lattices \cite{deis10,schr15,luki19,aban19}, posing a major obstacle for understanding many-body effects in dynamical localization \cite{zhan04,piko08,flac09,lell20,nota20,ryla20,chic21,vuat21}. In fact, conflicting theoretical predictions exist: while mean-field calculations for interacting Bose-Einstein condensates (BEC) predict delocalization in momentum space with sub-diffusive character \cite{zhan04,lell20} (i.e., weaker-than-linear growth of system energy), the low-energy approximation based on Luttinger liquid theory of a kicked 1D Lieb-Liniger gas shows the persistence of dynamical localization \cite{ryla20}.  

Here we perform the first experimental study of many-body effects in dynamical localization of a QKR and report the observation of an interaction-driven transition between dynamically localized and delocalized states. In our periodically kicked one-dimensional bosonic system with contact interactions, the delocalization is manifest as a clear onset of sub-diffusive energy growth with kick number as the interaction is strengthened through tight transverse confinement. The sub-diffusive behavior persists over a range of interaction strengths and kick parameters. Our theoretical modeling with mean-field and Hartree-Fock-Bogoliubov approaches reasonably capture the observed dynamics in the deep delocalization and localization regions. However, the mean-field theory fails across the phase transition boundary, potentially due to the strong competition between the disorder potential and interaction-induced infinite long-range coupling in the momentum space lattice, which is extremely challenging to model in theory.

\begin{figure}
		\center
		\includegraphics[width=0.45\textwidth]{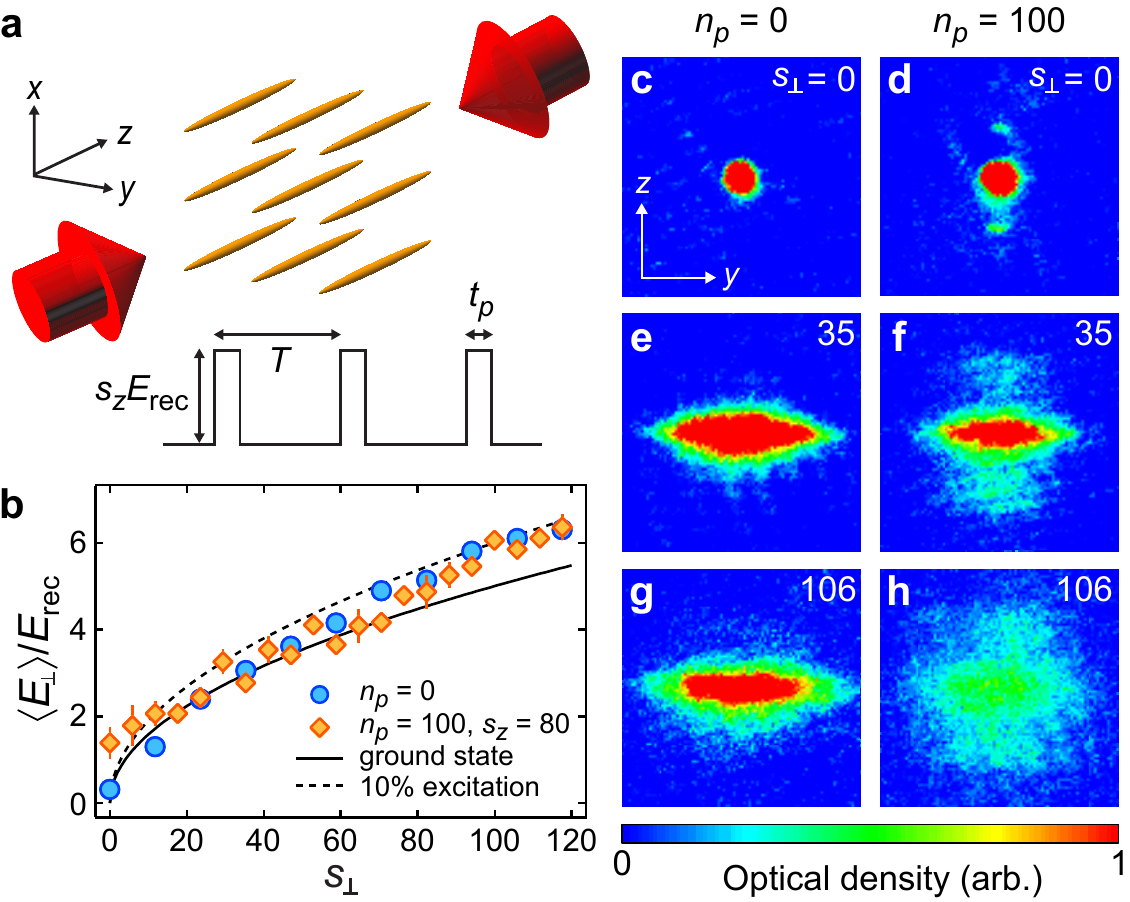}
		\caption{{\bf Experimentally realizing the interacting 1D quantum kicked rotor system.} ({\bf a}) Experiment schematic showing BECs in 1D tubes with periodic kicking pulses applied along the axial ($z$) direction. ({\bf b}) Average transverse energy $\langle E_{\perp} \rangle$ for various $s_{\perp}$ for no kick (circles) and 100 kicks with $s_z=80$ and $(t_p,T)=(2,105)\,\mu$s (diamonds). The solid line indicates the calculated energy for the transverse ground state and the dashed line that for 10\%  occupation of the first transverse excited state.({\bf c-h}) Time-of-flight atom absorption images after 0 (left column) and 100 (right column) kicks for $s_{\perp}=0$ (3D case) and $s_{\perp}=35$ and 106. For (d,f,h), $s_z=80$ and $(t_p,T)=(2,105)\,\mu$s. The imaging axis is along the $x$-direction and each image spans the momentum range $10 \; \hbar k_L \times 10 \; \hbar k_L$.}
		\label{fig:qkr_fig1}
\end{figure}

We initiate our experiments \cite{SUPP} by preparing an essentially pure 3D BEC containing $1.5\times 10^5$ atoms of $^{174}$Yb with chemical potential $h\!\times\!1.1\,$kHz in an optical dipole trap (ODT), and subsequently loading it into a two-dimensional optical lattice where the atoms reside in a set of 1D tubes with negligible inter-tube tunneling (Fig.~\ref{fig:qkr_fig1}a). The starting BEC fraction in the tubes is higher than $85\%$. The kicks are implemented by a pulsed one-dimensional optical lattice along the axial direction of the tubes. The three orthogonal lattices are each formed from retro-reflected laser beams ($\simeq 100\,\mu$m waist) and have spatial period $1073\,{\rm nm}/2\!=\!\pi/k_L$, with corresponding recoil energy $E_{\rm rec}\!=\!\hbar \omega_{\rm rec}$ where $\omega_{\rm rec}\!=\hbar k_L^2/2m\!=\!2\pi \times 1\,$kHz and $m$ is the atom mass. The kick parameters are tunable through the kick period $T$, pulse width $t_p$, and potential depth $s_z E_{\rm rec}$. Each of the two transverse lattices has depth $s_{\perp} E_{\rm rec}$. For a typical $s_{\perp}=106$ used in this work, the transverse trap frequency (for the central tube) is $\omega_{\perp}=2\sqrt{s_{\perp}} \;\omega_{\rm rec} = 2\pi\times20.5\,$kHz. The transverse oscillator length is $a_{\perp} = \sqrt{\hbar/m\omega_{\perp}} \simeq 53\,$nm, and the axial frequency is  $\omega_z = 2\pi \times 71\,$Hz. From the Thomas-Fermi (TF) radii of the 3D trap and the measured axial size in the 1D tubes, we estimate a particle number of $N_{\rm atom}=600$ per tube and an initial 1D peak density of $\bar{n}_{\rm 1D} = 25/\,\mu$m for the central tube.

We monitor the system by diabatically turning off all optical potentials after a desired number ($n_p$) of kicks and then taking a time-of-flight absorption image from which we extract the atomic momentum distribution in both axial and transverse directions. The measured transverse distribution is consistent with the transverse ground state energy as shown in Fig.~\ref{fig:qkr_fig1}b. The 1D geometry with $\omega_{\perp} \gg \omega_{\rm rec}$ suppresses two-body scattering from the axial to the transverse direction, as evident in the negligible growth of transverse energy $\langle E_{\perp} \rangle$ during the kicking process (Fig.~\ref{fig:qkr_fig1}b) for $s_{\perp}\geq 20$. As interactions are increased by raising $s_{\perp}$, the axial ($z$) momentum width after many pulses also increases ($n_p=0,100$ shown in Fig.~\ref{fig:qkr_fig1}c-h), providing a key signature for examining the many-body QKR.

\begin{figure}
		\center
		\includegraphics[width=0.47\textwidth]{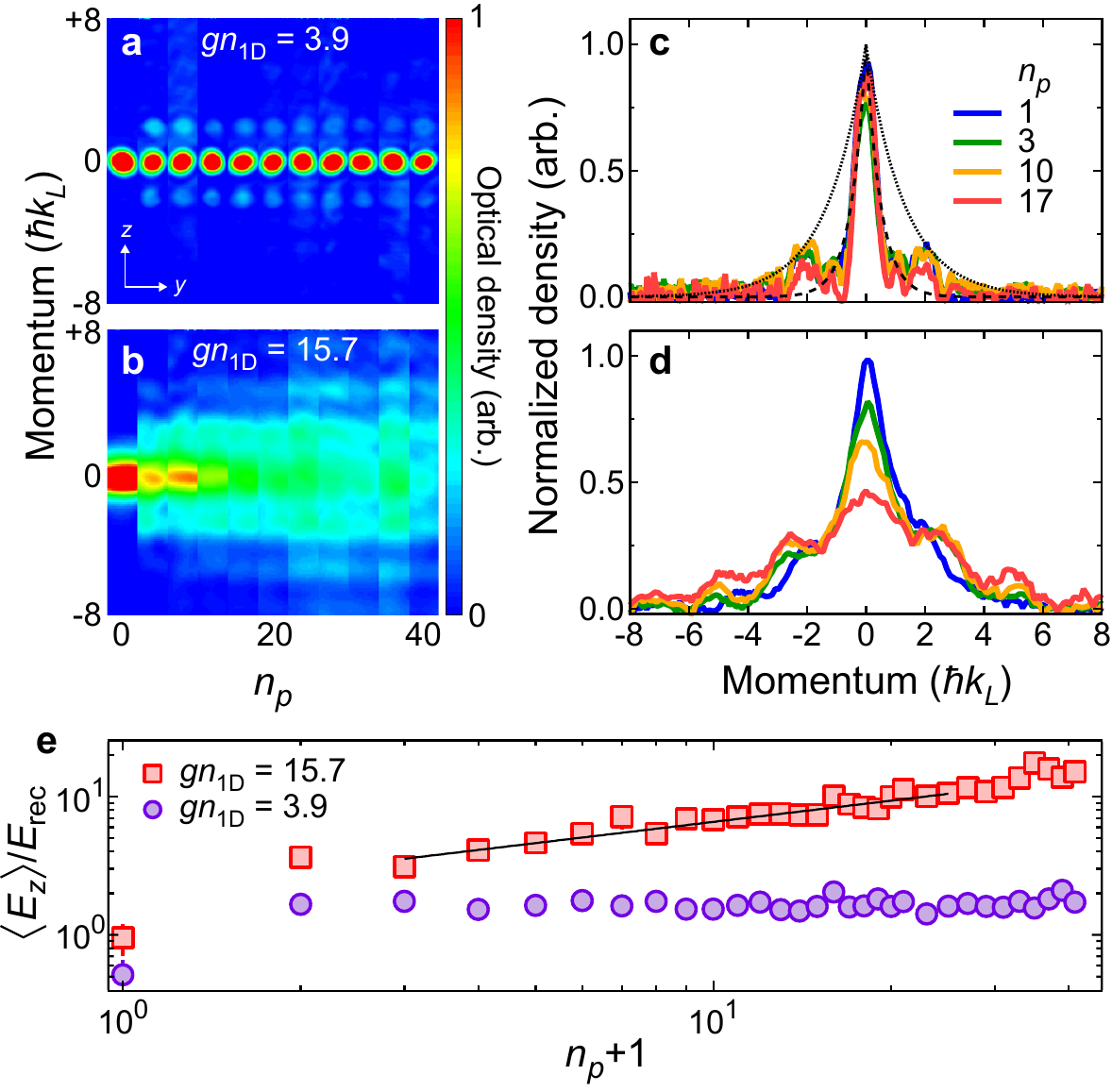}
		\caption{{\bf Momentum and energy evolution for dynamically localized and delocalized rotors.} ({\bf a-b}) Sequences of absorption images for (a) localized and (b) delocalized cases with $\kbar = 5.26$, $K=5.3$, and $(t_p,T)=(2,105)\;\mu$s. ({\bf c-d}) Axial momentum distributions after kick numbers $n_p=1,3,10,17$, corresponding to (a) and (b) respectively. The dashed and dotted lines in (c) are exponential functions (see text). ({\bf e}) Evolution of axial kinetic energy corresponding to (a) and (b). The solid line is a power law fit to the delocalized data returning an exponent value of 0.5.}
		\label{fig:qkr_fig2}
\end{figure}

Even though condensation is not possible in the homogeneous 1D case \cite{lieb63,lieb63b,yang69}, axial harmonic confinement supports BEC \cite{kett96}. For our experimental parameters, the system is quasi-1D where the gas is kinematically 1D with the two-body scattering length $a_s = 5.55\,{\rm nm} (\ll\!a_{\perp}$) retaining its 3D value. The correlation length $l_c=\hbar/\sqrt{m\bar{g}\bar{n}_{\rm 1D}}$ is much larger than the mean inter-particle separation $1/\bar{n}_{\rm 1D}$, which makes the ground state of the initial system a true TF condensate \cite{petr00}. Here $\bar{g}=2\hbar^2 a_s /(ma_{\perp}^2)$ is the mean-field interaction constant.

We model the many-body dynamics of bosons with the mean-field theory, where the QKR wavefunction $\Phi$ is governed by the non-linear Gross-Pitaevskii (GP) equation:
\begin{multline}
i\kbar \partial_\tau \Phi(\theta,\tau) =  \Bigg( \Bigg. -\frac{\kbar^2}{2} \partial^2_\theta - K \cos \theta \sum_{n_p} \delta(\tau-n_p) \\+ \frac{1}{2}{\omega}_\theta^2\theta^2 + g |\Phi(\theta,\tau)|^2 \Bigg) \Bigg. \Phi(\theta,\tau)
\label{eq:GP}
\end{multline}
with the dimensionless parameters $\theta = 2k_L z$, 
$\tau=t/T$, where $\kbar = 8\omega_{\rm rec} T$ is the dimensionless effective Planck constant. The dimensionless kick strength $K$ and interaction strength $g$ are defined as
\begin{equation}
    K = 4s_z \omega_{\rm rec}^2 t_p T \, ,\quad g=\frac{2\bar{g} k_L \kbar T}{\hbar} = \kbar^2 \frac{k_L a_s}{(k_L a_\perp)^2} \, .
    \label{eq:params}
\end{equation}
The dimensionless axial frequency is ${\omega}_\theta=\omega_zT$ and the dimensionless initial peak density is $n_\text{1D}=|\Phi(0,0)|^2 = {\bar n}_{\rm 1D}/2k_L$, where the wave function is normalized as $\int d\theta |\Phi(\theta,\tau)|^2 =N_\text{atom}$. Throughout this paper, we label the interaction strength $gn_\text{1D}$ using an average value that takes into account the variation of atom number in different tubes. For the 3D case, $gn_{\rm 1D}=3.9$ is obtained by adjusting $a_{\perp}$ in Eq. 2 to match the measured chemical potential, which corresponds to $s_\perp=1.6$.

\begin{figure}
		\center
		\includegraphics[width=0.46\textwidth]{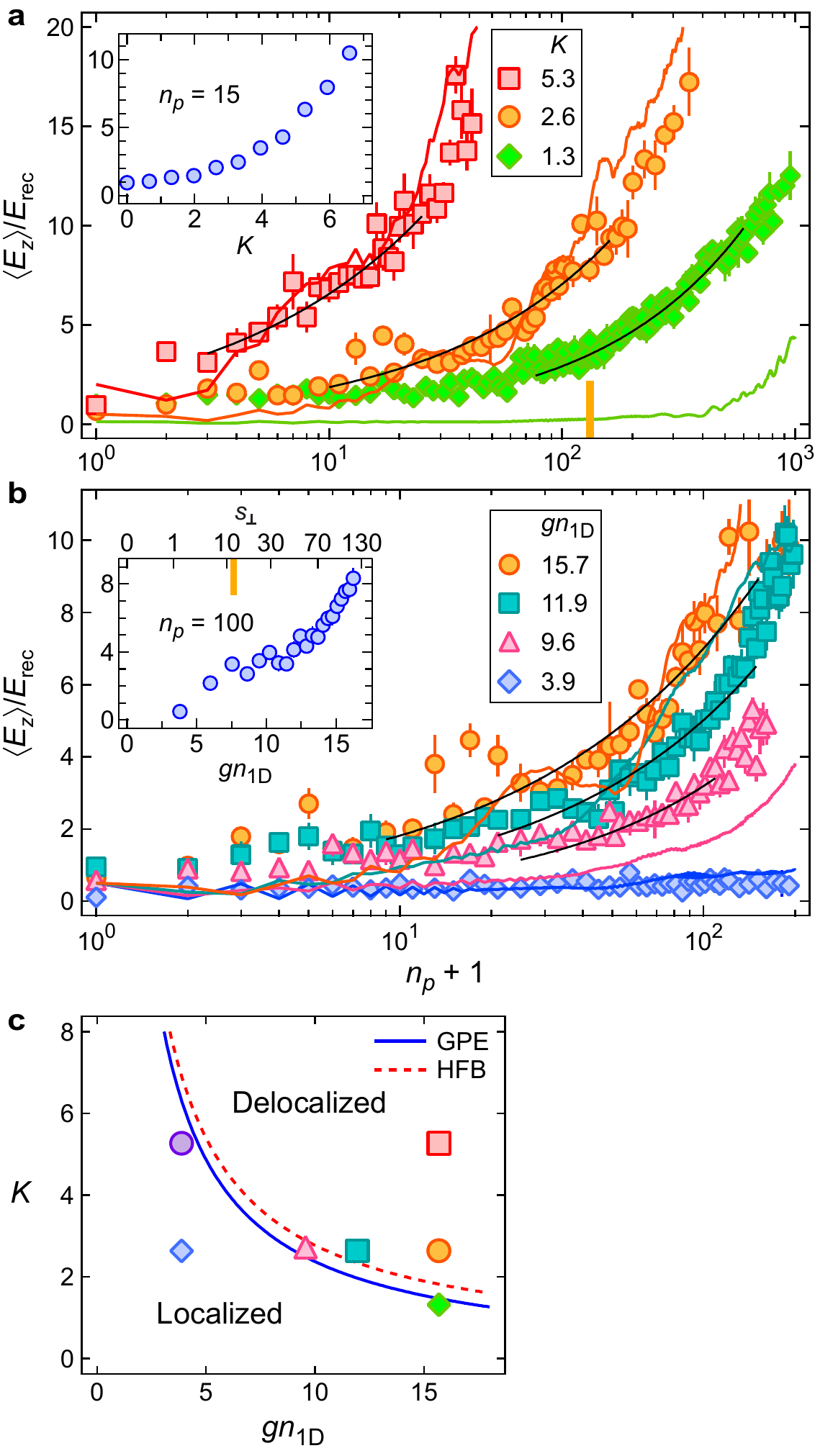}
		\caption{{\bf Tuning the onset of many-body quantum chaos with interaction and kick strengths.} For all data in this figure, $(t_p,T) = (2,105)\,\mu s$, and $\kbar=5.26$. ({\bf a}) Evolution of $\Ez$ with pulse number for three different kick strengths with $gn_{\rm 1D}=15.7$. The yellow vertical bar indicates the axial oscillation period. The inset shows $\Ez$ after 15 kicks for various $K$. ({\bf b}) Evolution of $\Ez$ with pulse number for four different interaction strengths with $K=2.6$. The inset shows $\Ez$ after 100 kicks for various $gn_{\rm 1D}$, with the yellow vertical bar marking inter-tube tunneling time of $21\,$ms ($n_p=200$). To preserve a similar axial trapping frequency for low $s_\perp$, the ODT is kept on for the data in (b) and its inset, except for the $gn_\text{1D} = 15.7$ data (orange circles). The colored solid lines in (a) and (b) are corresponding numerical simulations using the GP equation. The black solid lines in (a) and (b) are power law fits to the delocalized data which return exponent values of \{0.51, 0.58, 0.68\}, for $K=\{5.3, 2.6, 1.3\}$ in (a) and \{0.58, 0.65, 0.74\} for $gn_{\rm 1D}=\{15.7, 11.9, 9.6\}$ in (b). The statistical errors on the fitted exponents are $ \leq 8\%$.  ({\bf c}) Phase diagram of the localization-delocalization behavior of the system for $\kbar=5.26$, with lines indicating the phase boundaries by solving the GP (solid) and HFB (dashed) equations. The filled markers correspond to the data from (a,b) and Fig.~\ref{fig:qkr_fig2}e.}
		\label{fig:qkr_fig3}
\end{figure}

We first discuss QKR experiments on a 3D system, implemented by applying the pulsed lattice on the BEC trapped in the ODT with no transverse lattice. Here we always observe dynamical localization as shown in Fig.\,\ref{fig:qkr_fig2}a,c,e, consistent with weak interactions. Following some initial coherent dynamics, the momentum distribution and mean energy $\Ez$ quickly saturate. Compared to the exponentially localized function $e^{-|p|/\xi}$, the observed momentum profile exceeds the expected dynamical localization length $2\xi = K^2/2\kbar^2 = 0.5$ (in units of $\hbar k_L$) for a non-interacting system, but is better contained by $2\xi= 1.3$ (see dashed and dotted lines in Fig.~\ref{fig:qkr_fig2}c), corresponding to the observed saturated value $\Ez = 1.7\,E_{\rm rec}$. The deviation from the non-interacting prediction reflects the small but non-zero $g$. In striking contrast to the 3D case, many-body dynamical delocalization is evident for the higher interaction strengths available in the 1D geometry, with a sub-diffusive energy growth, as displayed in Fig.~\ref{fig:qkr_fig2}b,d,e.

Figure~\ref{fig:qkr_fig3} shows a study of the delocalization behavior for different kick strengths $K$ (tuned through $s_z$) and interaction strengths (tuned through $s_\perp$). The dynamics are strongly dependent on $K$, as seen in Fig.~\ref{fig:qkr_fig3}a, with the earliest onset of delocalization for $(gn_{\rm 1D},K)=(15.7,5.3)$ exhibiting significant energy growth even within the first 10 pulses. We note that this timescale ($1\,$ms) is more than ten times shorter than the axial oscillation period, indicating that harmonic confinement along the tube is not a pre-requisite for the observed delocalization. As $K$ is lowered, delocalization is delayed and a prethermal plateau emerges prior to monotonic energy growth. For $\Ez$ above $10\,E_{\rm rec}$ we also observe number loss as the most energetic atoms are ejected out of the kicked system \cite{SUPP}. Power law fits of the form $E_0 t^{\alpha}$ to $\Ez$ growth below $10\,E_{\rm rec}$ yield $\alpha$ between 0.5 and 0.7. The observed sub-diffusivity is intermediate between classical chaotic behavior and single particle quantum mechanics, and signals the fundamentally distinct dynamics of a driven quantum many-body system.

We investigate the dependence of delocalization behavior on interaction strength by changing the external confinement through $s_{\perp}$, which changes $a_\perp$ and hence $g$ and $n_{\rm 1D}$. As shown in Fig.~\ref{fig:qkr_fig3}b, we find stronger delocalization with higher $gn_{\rm 1D}$ for fixed $K=2.6$, with only the lowest $gn_{\rm 1D}=3.9$ case remaining localized.

We carry out numerical mean-field simulations of the dynamics (solid lines in Fig.~\ref{fig:qkr_fig3}a,b), starting from the TF ground state obtained by the imaginary time evolution of the GP equation \cite{SUPP}. We find that for a given interaction strength $gn_{\rm 1D}$, the system enters a dynamically delocalized phase with mean energy increasing with pulse number when $K$ is larger than a critical value $K_c$. As $gn_{\rm 1D}$ is increased, $K_c$ decreases, implying that dynamical delocalization is easier for stronger interactions. The variation of $K_c$ marks the boundary (solid line in Fig.~\ref{fig:qkr_fig3}c) between localized and delocalized phases in the $K$-$gn_\text{1D}$ parameter space. We see reasonable agreement between theory and experiment in the system time evolution for points deep in the delocalization and localization regimes. However, significant deviations between the two exist for points near the phase boundary. 

The delocalization physics may be understood more intuitively in momentum space, where the non-interacting QKR can be mapped to a one-dimensional lattice with on-site disorders~\cite{fish82}. In momentum space, the real-space contact interaction introduces on-site nonlinear terms as well as infinite long-range nonlinear cross-hopping terms to the disordered lattice, which are responsible for the dynamical delocalization \cite{SUPP}. The observed failure of the mean-field approach near the phase boundary is therefore unsurprising, since the dynamics near $K_c$ are very sensitive to the competition between the disorder potential and the interaction-induced infinite long-range coupling in the momentum space lattice. We also note that the finite momentum width of the initial state and the interaction-induced scattering between different momenta may lead to density peaks away from the recoil momenta ($2jk_L$ with integer $j$), which can happen at a small kick number for large $K$ (see Fig.~\ref{fig:qkr_fig2}).

Initially our system is a true TF condensate with negligible fluctuations. As the kick number increases, the delocalization of the interacting system is accompanied by a rapid proliferation of non-condensate particles. The mean-field GP approach is valid only when quantum depletion is low, i.e., the non-condensate fraction is much smaller than the condensed fraction. Going beyond mean-field, we examine the excitation properties by employing the Hartree-Fock-Bogoliubov (HFB) \cite{grif96, zhan04, SUPP} approximation to calculate the evolution of the non-condensate particle number $\langle\hat{\psi}^\dag\hat{\psi}\rangle$, where $\hat{\psi}$ represents the quantum fluctuation beyond the condensate $\Phi(\theta,\tau)$.
The dashed line in Fig.~\ref{fig:qkr_fig3}c represents the boundary between stable and unstable regimes, where the unstable regime is manifest as an exponential increase of non-condensate particles with $n_p$. The two phase boundaries (solid and dashed lines in Fig.~\ref{fig:qkr_fig3}c) are close to each other, suggesting that the dynamical delocalization is accompanied by the BEC instability. 

\begin{figure}
		\center
		\includegraphics[width=0.45\textwidth]{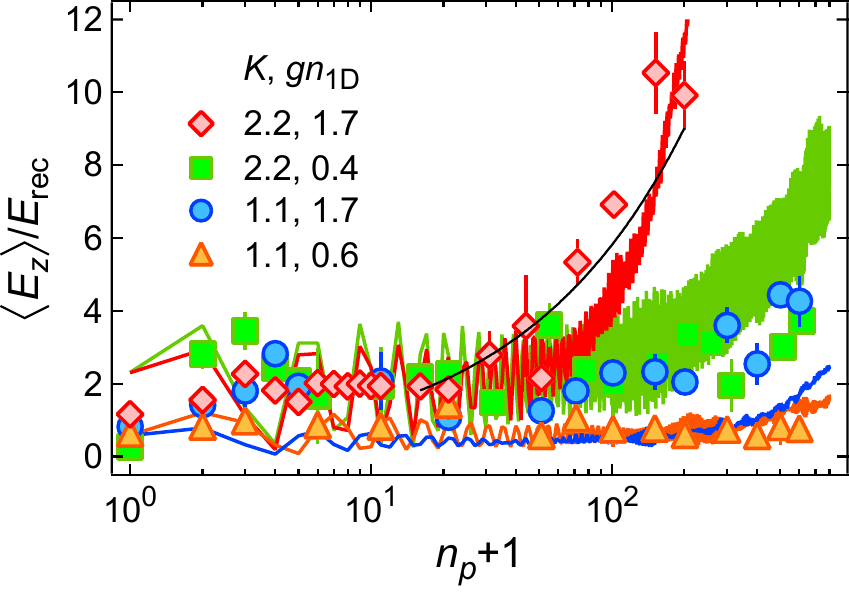}
		\caption{{\bf Tuning delocalization dynamics with number variation.} Shown are the evolution of mean energy for different number densities and kick strengths with $(t_p,T)=(3.5,40)\,\mu$s, $\kbar=2.0$ in 1D tubes with $s_{\perp}=106$, $g=1.3$, and axial frequency $2\pi \times 161\,$Hz \cite{SUPP}. The colored solid lines are the corresponding numerical simulations using the GP equation and the black solid line is a power law fit to the delocalized data returning an exponent value of 0.63.}
		\label{fig:qkr_fig4}
\end{figure}

In addition to the transverse confinement, the interaction strength in 1D tubes can also be tuned using the atom number $N_{\rm atom}$ (thus $n_{\rm 1D}$). As shown in Fig.~\ref{fig:qkr_fig4}, we observe that the QKR behavior changes dramatically with change of initial $n_{\rm 1D}$ but keeping the same $g$. As before, the numerical GP solutions track our experimental observations reasonably well when far away from the delocalization-localization boundary, but show significant deviations near it. The larger amplitude oscillations in the numerics are a consequence of the shorter period in the kick pulse parameters $(t_p,T)=(3.5,40)\,\mu$s compared to that in Fig.~\ref{fig:qkr_fig3}. The existence of a prethermal localized plateau followed by many-body delocalization for different periods highlights the generality of the observed phenomenon. Indeed we have observed such many-body effects over a wide range of kick periods \cite{SUPP}.

Our results experimentally realize the interacting QKR, a long-sought quantum mechanical paradigm system. The combined experimental and theoretical study of many-body localized and delocalized phases in momentum space spotlight the emergence of many-body quantum chaos and constitute the first study of the effects of interactions on dynamical localization, an area where current theoretical results are in conflict \cite{zhan04,lell20,ryla20}. Direct extensions of these studies include further characterization of the boundary between localized and delocalized phases where we observe the mean-field theory to fail, the nature of the prethermal state, and the possible universality \cite{flac09,lell20} of the sub-diffusive delocalization exponent. It will also be interesting to extend the current implementation carried out with $\gamma = 1/(l_c \bar{n}_{\rm 1D})^2 \ll 1$ into the $\gamma \gg 1$ Tonks-Girardeau regime \cite{pare04,kino04} where beyond mean-field theories predict many-body dynamical localization with momentum profile distinct from their spatially localized counterpart \cite{ryla20,chic21,vuat21}. Our technique of tight confinement to tune interactions in the synthetic momentum space can also be extended towards studies of the momentum-space Josephson effect \cite{houj18}, interaction-driven transport in higher synthetic dimensions \cite{cher14,verm20}, and topological phases with interactions in coupled momentum space lattices \cite{meie18}. 

{\it Acknowledgments}
We thank David Weld, Adam Rançon, and Victor Galitski for helpful discussions. During the course of this work, we became aware of related advances in another experimental group \cite{caoa21}. {\bf Funding:} The work at University of Washington is supported by the Air Force Office of Scientific Research (FA9550-19-1-0012) and the National Science Foundation (PHY-1806212). K.C.M. is supported by an IC postdoctoral fellowship. The work at the University of Texas at Dallas is supported by the Air Force Office of Scientific Research (FA9550-20-1-0220), National Science Foundation (PHY-1806227, PHY-2110212), and Army Research Office (W911NF-17-1-0128). 
\\ \\
{\bf Methods}\\ \\
\emph{Experimental Setup}\\
The experiments discussed in this work were performed in an apparatus discussed in earlier work \cite{hans11,roy16,roy17} and augmented with a set of three mutually orthogonal and independently controlled optical lattices. We prepare a BEC containing $1.5\times 10^5$ atoms of $^{174}$Yb atoms in a crossed optical dipole trap (ODT) \cite{roy16} with trapping frequencies $\{\omega_{0x},\omega_{0y},\omega_{0z}\} = 2\pi \times \{145,16,53\}$ Hz, chemical potential $h \times 1.1\,$kHz, and corresponding Thomas-Fermi radii of $\{2.4,22,6.6\}\,\mu$m. The BEC is then transferred into a two-dimensional optical lattice formed by two pairs of counter-propagating laser beams where the atoms reside in a set of 1D tubes (see Fig.~1a in main text). The lattice and kick optical potentials are derived from a home-built external cavity diode laser operated at wavelength $\lambda=1073\,$nm, and amplified by a 50W amplifier (NuFern NUA-1064-PD-0050-D0). The total laser power for the lattice beams is distributed between three paths -- two for the two-dimensional transverse optical lattice and one for the axial kicking lattice. To suppress optical interference between different paths from affecting the atoms, for each pair of paths, we maintain orthogonal linear polarizations and use acousto-optic modulators (AOMs) to establish $>40\,$MHz frequency separation. The laser beams forming the two-dimensional transverse lattice are intensity stabilized at the 2\% level. We calibrate the depth of our lattices using single-pulse Kapitza-Dirac diffraction, a procedure which also provides an experimental measurement of our lattice beam waists to be $\{w_x,w_y,w_z\}=\{121,101,99\}\,\mu$m, which are much larger than the BEC size. 
\\ \\
\emph{Loading and characterizing the 1D gas} \\
To transfer the BEC from the ODT to 1D confinement, the transverse lattice is ramped up exponentially in $100\,$ms with an exponential time constant of $20\,$ms. To assess the adiabaticity of this process, we have performed tests in which the forward (loading) ramp is immediately followed by a reverse ramp back to the ODT after which we compare the final BEC fraction to the initial value. Starting from an essentially pure BEC, we obtain about 70\% BEC fraction after the forward and reverse ramps, suggesting that the BEC fraction is about 85\% in the two-dimensional lattice. We believe this number to be a lower bound because the recovered BEC fraction is likely also limited by the lack of coherence between the tubes, as tunneling is strongly suppressed beyond $s_{\perp} \simeq 20$.

We load $1.2\times10^5$ atoms into about 180 horizontal tubes, as determined by the initial Thomas-Fermi radii in the 3D trap. We measure an initial (tube-averaged) axial size of $27\,\mu$m for $s_{\perp}=106$. The peak density of the central tubes is then ${\bar n}_{\rm 1D}=25\,/\mu$m. Once the BEC is loaded into the two-dimensional lattice, we exponentially ramp down the ODT in $50\,$ms with a time constant of $10\,$ms, before pulsing on the kick laser along the axial direction of the tubes. To obtain the momentum distribution, we diabatically turn off all the optical potentials and take an absorption image of the atoms after a long time-of-flight (TOF) set to a value between 15 and $43\,$ms. 

The observed growth rate of the axial momentum distribution in the absence of kicking pulses determines the background heating rate in the system. For $s_{\perp}=106$, we measure a kinetic energy growth rate of $6\,E_{\rm rec}$/s. All of the QKR experiments reported in this work occur within $100\,$ms, a timescale during which this background heating is negligible. The calculated photon scattering rate from the transverse lattice is $<0.1$/sec for $s_{\perp}=106$, suggesting that the observed residual heating is from technical noise.
\\ \\
\emph{Kicking Pulse Implementation}\\ 
The kick pulses are generated by triggering a function generator (Stanford Research Systems DS345) to produce a desired sequence of TTL pulses, which in turn controls the radio-frequency switch driving the AOM for the kicking lattice laser beam. By integrating over the observed axial momentum distribution of the atoms in the time-of-flight absorption image, we calculate the kinetic energy delivered to the system by the kicks. 
\\ \\
\emph{Nonlinear Anderson model in the momentum space lattice---}The QKR is a
Floquet system with the wave function $\Phi (\theta,\tau)=e^{-i\epsilon \tau}\phi (\theta,\tau)$%
, where $\phi (\theta,\tau)=\phi (\theta,\tau+1)$ is the periodic part and $\epsilon$ is the quasienergy. The $\delta $-kick
leads to $\phi _{+}(\theta)=e^{iK\cos(\theta)/\kbar}\phi _{-}(\theta)$, where $%
\phi _{\pm }(\theta)$ are the wave functions just after and before the kick. In
the momentum space lattice, the free evolution between kicks is described by 
\begin{eqnarray}
i\kbar \partial _{\tau}\phi _{j}(\tau) &=&\left( \frac{\kbar^{2}j^{2}}{%
2}-\kbar \epsilon \right) \phi _{j}(\tau)  \notag \\
&+&\frac{g}{4\pi M}\left( 2 N_{\text{atom}}-|\phi _{j}|^{2}\right) \phi _{j}(\tau)  \notag
\\
&+&\frac{g}{4\pi M}\left[ \sum_{j^{\prime }\neq 0,j_{1}-j}\phi _{j_{1}}^{\ast }\phi
_{j_{1}-j^{\prime }}\right] \phi _{j+j^{\prime }}(\tau),
\label{Momentumlattice}
\end{eqnarray}%
where $\phi _{j}(\tau)$ is the $j$-th Fourier component of the wave function $\phi
(\theta,\tau)$ (i.e., at the momentum site $j$) and $M=Z k_L/2 \pi$ with the system size $Z$. The first line on the right-hand side corresponds to the single-particle evolution, which leads to momentum space dynamical localization in the non-interacting QKR. The second and third lines correspond to diagonal (on-site attraction) and off-diagonal (infinite long-range hopping) interactions.

If only the diagonal interaction is considered, the free evolution yields $%
\phi_{-,j}=\phi _{+,j}\exp(i[\epsilon - \kbar j^{2}/2 - {g}(2N_{\text{%
atom}}-|\phi _{+,j}|^2)/4\pi M\kbar ])$ and the nonlinear Anderson model becomes 
\begin{equation}
V_{j}\bar{\phi}_{j}+\sum_{j^{\prime }\neq 0}K_{j^{\prime }}\bar{\phi}%
_{j+j^{\prime }}=\omega \bar{\phi}_{j},
\end{equation}%
which has the same form as that for the non-interacting QKR except that the on-site
disorder is nonlinear with $V_{j}=\tan [\epsilon /2- \kbar j^2/4
 -{g}N_{\text{atom}}/4\pi M\kbar + {g}|\sum_{j^{\prime }}\bar{%
\phi}_{j+j^{\prime }}(K_{j^{\prime }}+\delta _{j,j^{\prime }})|^{2}/8\pi M\kbar ]
$. Here $\bar{\phi}_{j}=({\phi }_{-,j}+{\phi }_{+,j})/2$, hopping rates $%
K_{j}=\frac{1}{\sqrt{4M\pi}}\int d\theta e^{ij\theta}\tan [\frac{{K }}{2\kbar}\cos
(\theta)]$, and energy $\omega =-K_{0}$. In the presence of infinite
long-range hopping in the momentum space, the dynamics are much more complex,
without an explicit relation between $\phi _{-}$ and $\phi _{+}$. Such infinite
long-range hopping destroys the quantum interference in the momentum space,
leading to dynamical delocalization.

\bibliography{mixrefs21}

\clearpage
\onecolumngrid
\renewcommand{\theequation}{S\arabic{equation}}
\renewcommand{\thefigure}{S\arabic{figure}}
\setcounter{equation}{0}
\setcounter{figure}{0}
\begin{center}
    {\Large \bf Supplementary Materials for ``Observation of Many-body Dynamical Delocalization in a Kicked 1D Ultracold Gas"}
\end{center}

\section{Experimental Details}

\emph{Optical Lattice Implementations.---} The 2D lattice potential is always applied such that each of the two arms contributes a peak (AC Stark) potential depth $V_{\rm lat}$. The overall lattice potential is then:
\begin{equation}
-V_{\rm lat}\left[\frac{e^{-\frac{2(y^2+z^2)}{w_x^2(1+x^2/x_R^2)}}}{1+x^2/x_R^2} \cos^2 (k_Lx) + \frac{e^{-\frac{2(x^2+z^2)}{w_y^2(1+y^2/y_R^2)}}}{1+y^2/y_R^2} \cos^2 (k_Ly) \right]
\end{equation}
where $j_R=\pi w_j^2/\lambda$ is the Rayleigh range along the $j^{\rm th}$ direction ($j$ is $x$ or $y$). The full depth of the lattice is $2V_{\rm lat}$. By taking the lowest order expansion of this expression in $x,y,z$, we obtain the radial and axial trap frequencies for the center tube located at the origin:
\begin{equation}
    \omega_{j} = \sqrt{\frac{2V_{\rm lat}}{m} \left( \frac{2}{w_{j}^2}+\frac{1}{j_R^2}+k_L^2 \right) } \simeq \sqrt{\frac{2V_{\rm lat} k_L^2}{m} } = 2 \omega_\text{rec} \sqrt{s_{\perp}} = \omega_{\perp}, 
\quad
    \omega_{z} = \sqrt{\frac{4V_{\rm lat}}{m}(\frac{1}{w_x^2}+\frac{1}{w_y^2})} = 2 \frac{\lambda}{\pi \bar{w}} \omega_\text{rec} \sqrt{s_{\perp}} 
\end{equation}
where $2/{\bar w}^2=1/w_x^2+1/w_y^2$, $s_{\perp}=V_{\rm lat}/E_{\rm rec}$, $E_{\rm rec} = \frac{\hbar^2 k_L^2}{2m}$ is the recoil energy and $m$ is the mass of an $^{174}$Yb atom. Since the lattice beam waists are much larger than the initial BEC size, all tubes containing atoms experience approximately the same trap frequencies. 

\begin{figure}[ht]
		\center
		\includegraphics[width=0.4\textwidth]{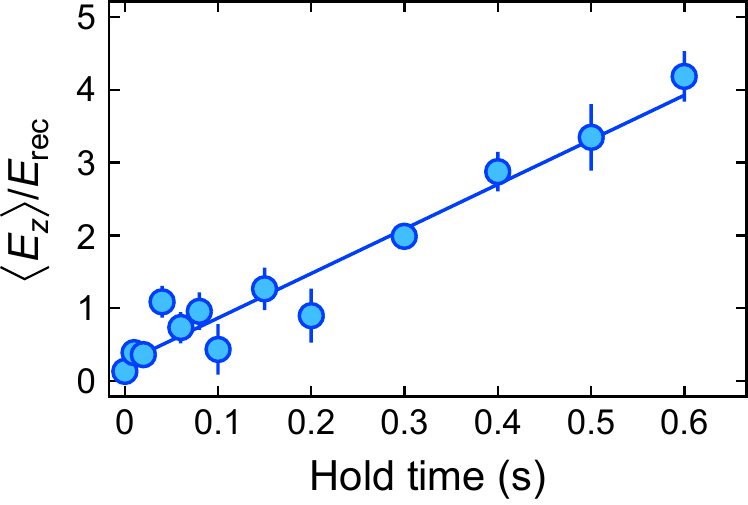}
		\caption{{\bf Background heating rate in lattice.} The plot shows the measured mean energy in the absence of pulses, after different hold times in a two-dimensional lattice of depth $s_\perp = 106$. Also shown is the linear fit, which gives a background heating rate of $6\,E_{\rm rec}$/s, which is negligible on the timescale of the QKR experiments.}
		\label{fig:figqkrsupp1}
\end{figure}

\begin{figure}[ht]
		\center
		\includegraphics[width=0.6\textwidth]{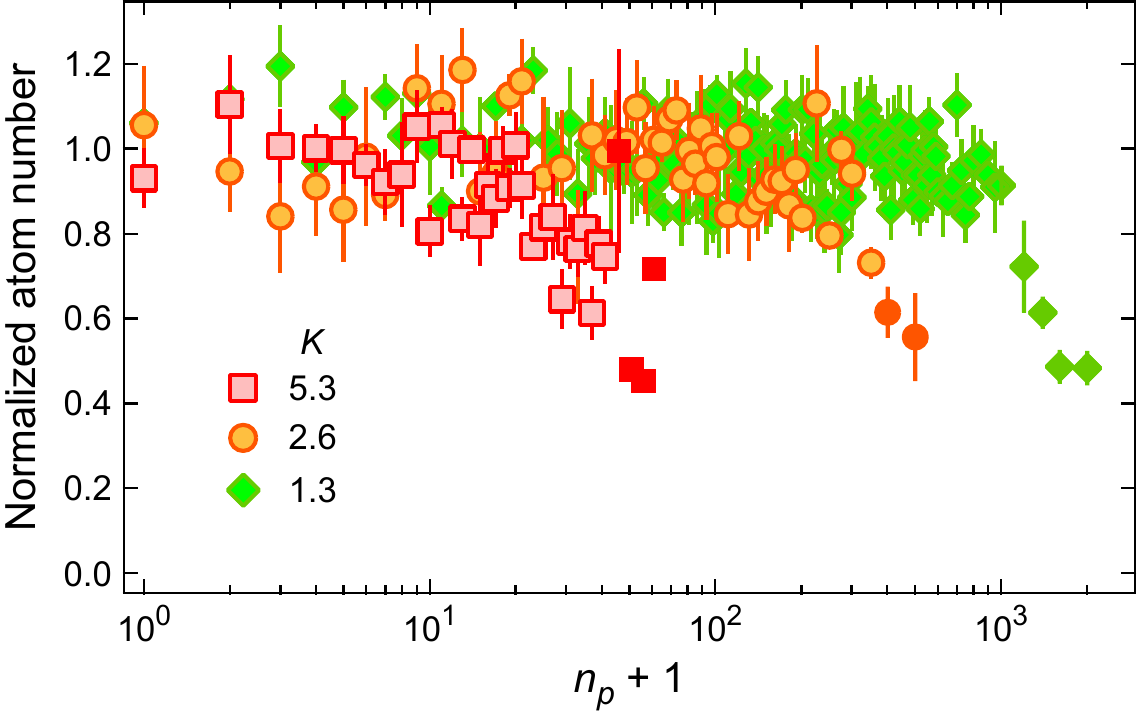}
		\caption{{\bf Atom number evolution during kicking process.} Normalized atom number as a function of pulse number $n_p$ for the data corresponding to the mean energy evolution presented in Fig. 3a in the main text. For each $K$, the darker colored markers (with the same shape) correspond to the largest pulse numbers that are not included in Fig. 3a of the main text.}
		\label{fig:figqkrsupp2}
\end{figure}

We load $1.2\times10^5$ atoms into about 180 horizontal tubes, as determined by the initial Thomas-Fermi radii in the 3D trap. The separation of individual tubes is below our imaging resolution, preventing access to accurate individual in-trap 1D TF profiles. We instead perform a Gaussian fit to the in-trap spatial distribution and measure an initial (tube-averaged) axial size of $27\,\mu$m for $s_{\perp}=106$. The peak density of the central tubes is then ${\bar n}_{\rm 1D}=25\,/\mu$m. 

By observing the growth in the axial momentum distribution as the atoms are held in the lattice in the absence of kicking pulses, we determine the background heating rate for atoms trapped in 1D tubes. As shown in Fig.~\ref{fig:figqkrsupp1}, for $s_{\perp}=106$, we measure a kinetic energy growth rate of $6\,E_{\rm rec}$/s.

The kicking beam waist and Rayleigh range are much larger than the initial BEC size and we can express the kicking pulse potential as $V_\text{kick} \cos^2 (k_L z) = s_z E_{\rm rec} \cos^2 (k_L z)$. As shown in Fig.~1b of the main text, even after 100 kicks with $s_z=80$, when the axial delocalization has become significant, the transverse excitation is negligible at $<10\%$. The 1D geometry also suppresses losses from three-body recombination \cite{tolr04}. Spontaneous scattering from lattice photons are also insignificant for our experimental parameters. Indeed atom loss in our 1D system only ensues when the QKR has reached a mean energy $\Ez$ of $10\,E_{\rm rec}$ for $s_\perp=106$, due to the finite trap depth. This can be seen in Fig.\,\ref{fig:figqkrsupp2}, in comparison with the delocalization data presented in Fig.~3a of the main text. \\

\begin{figure}[ht]
		\center
		\includegraphics[width=0.4\textwidth]{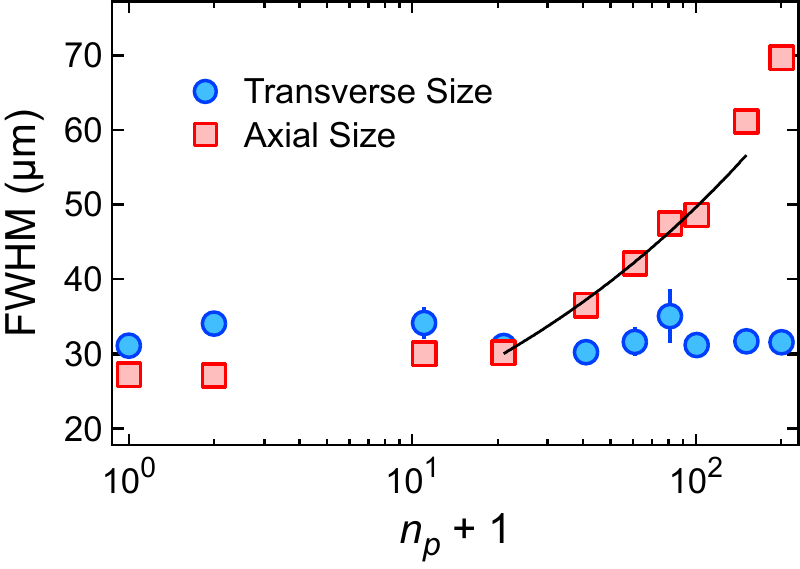}
		\caption{{\bf Evolution of cloud size during kicking process.} Axial ($z$) and transverse ($y$) size of the atom cloud confined in the $s_\perp=106$ lattice versus pulse number with $s_z = 80$, $T = 105 \mu$s, $K = 2.6$, $g n_\text{1D}=15.7$. The transverse size does not change over the course of the experiment, while the axial size starts to grow after the onset of dynamical delocalization. The solid line is a power law fit to the growth part and returns an exponent value consistent with that observed in momentum space.}
		\label{fig:figqkrsupp3}
\end{figure}

\emph{In-trap size during kicking.---} Since our QKR implementation is in a geometry that is not closed (such as a ring trap), we can expect that momentum diffusion will also lead to spatial diffusion. Shown in  Fig.~\ref{fig:figqkrsupp3} are measured in-trap axial ($z$) and transverse ($y$) sizes (full-width-half-max) for the conditions corresponding to the $K=2.6$ data in Fig. 3a of the main text, where mean energy evolution was presented. The measured in-trap axial size starts at $27\,\mu$m and indeed starts to grow after dynamical delocalization has set in (beyond 20 pulses). Fitting the growth part of Fig.\,\ref{fig:figqkrsupp3} with a power law function of the form $z_0 t^\beta$ returns $\beta=0.32(6)$. This would suggest a corresponding power-law growth of $2\beta$ for $\langle p^2 \rangle$ or $\Ez$ for these conditions, which is consistent with Fig.~3a of the main text. As expected, the transverse FWHM size does not change throughout the experiment and remains consistent with the initial TF radius in the $y$-direction.\\

\begin{figure}[ht]
		\center
		\includegraphics[width=0.5\textwidth]{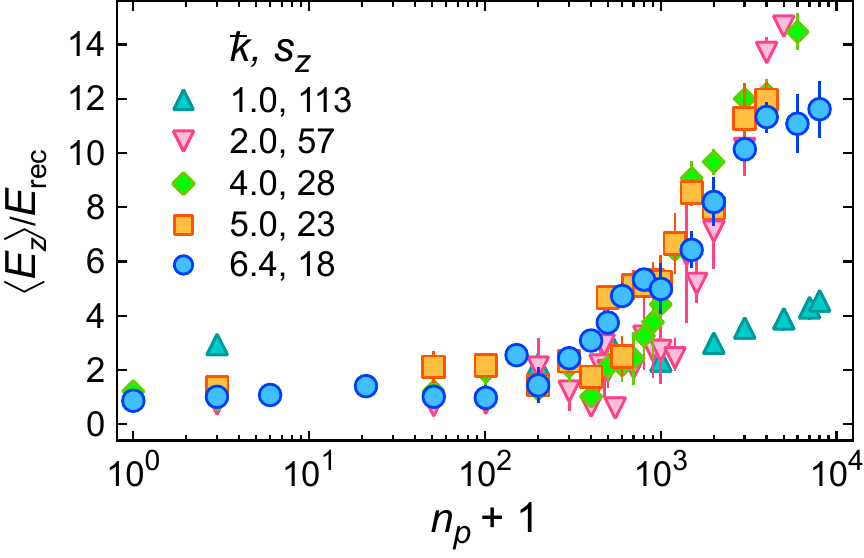}
		\caption{{\bf Onset of many-body dynamical delocalization for various kick periods.} Shown are mean energies as a function of pulse number for different kick periods $T$ (thus $\kbar$) and kick lattice depth $s_z$, while keeping the kick strength $K=0.7$ constant. The kick periods used are $T=\{20, 40, 80, 100, 125\}\mu $s, corresponding to $\kbar=\{1.0, 2.0, 4.0, 5.0, 6.4\}$.}
		\label{fig:figqkrsupp4}
\end{figure}

\emph{QKR with different pulse periods and orientation.---} Within the parameter space available in our experiment, we have observed dynamical delocalization over a large range of kick periods. Figure~\ref{fig:figqkrsupp4} shows the evolution of $\Ez$ for five period values in the range $T=20\,\mu$s to $125\,\mu$s, with $t_p=2\,\mu$s, while keeping the kick strength fixed at $K=0.7$. We also note that the period value of $T=125\,\mu$s corresponds to the quantum anti-resonance condition, where we would expect oscillatory behavior in the non-interacting case.

For all of the QKR experimental data presented in this work, the tubes are oriented along the horizontal $z$ direction, with the exception of the data shown in Fig. 4 of the main paper, where the tubes are oriented along the vertical $x$ direction. In order to counter gravity during the kick pulses, the ODT is not extinguished, leading to the higher axial frequency of $2\pi\times 161\,$Hz, the quadrature sum of the ODT and two-dimensional lattice contributions.    

\section{Theoretical Description}

\emph{Gross-Pitaevskii Mean-field approach.---}As discussed in the main
text, the transverse dynamics are suppressed by a strong 2D lattice, and the system contains many nearly decoupled quasi-1D tubes. The Hamiltonian of each tube reads 
\begin{equation}
\mathcal{H}=\int dz\hat{\Psi}(z)^{\dag }[H_{0}+\frac{\bar{g}}{2}\hat{\Psi}%
(z)^{\dag }\hat{\Psi}(z)]\hat{\Psi}(z)
\end{equation}%
with $\hat{\Psi}(z)$ the field operator of the Bose gas. The reduced 1D
interaction strength is $\bar{g}=\frac{2\hbar ^{2}a_{s}}{ma_{\perp }^{2}}$, and
the single-particle Hamiltonian is $H_{0}(t)=\frac{p^{2}}{2m}+V(z)-\hbar {%
\kappa }\cos (2k_{L}z)\sum_{n_{p}}\delta (t-n_{p}T)$, where $\hbar {k_{L}}$ is the
recoil momentum, and the kick strength $\hbar {\kappa }=V_{\text{kick}%
}t_{p}/2$. The dynamics of the nonlinear QKR are
described by 
\begin{equation}
i\hbar \partial _{t}\hat{\Psi}(z,t)=[H_{0}(t)+\bar{g}\hat{\Psi}(z,t)^{\dag }%
\hat{\Psi}(z,t)]\hat{\Psi}(z,t).  \label{eq:nlse}
\end{equation}

The regime of a Thomas Fermi (TF) condensate 
requires the correlation length $l_{c}=\frac{\hbar }{\sqrt{m\bar{g}\bar{n}%
_{\rm 1D}}}$ to be much larger than the mean inter-particle separation $1/\bar{n}%
_{\rm 1D}$, that is, the total atom number $N_{\text{atom}}$ in each tube should
be much larger than $N^{\ast }=(\frac{2a_{s}a_{z}}{a_{\perp }^{2}})^{2}$,
with $a_{z}=\sqrt{\frac{\hbar }{m\omega _{z}}}$. Even for strong
confinement along the transverse direction with $s_{\perp }=106$, $N^{\ast
}\simeq 15$. In our experiment, there are a few hundreds of atoms in each tube (%
$N_{\text{atom}}\gg N^{\ast }$), therefore, the system is in the weakly
interacting regime and we have a true TF condensate
at low temperature~\cite{petr00}. In the opposite limit with $N_{%
\text{atom}}\ll N^{\ast }$, the system would enter the Tonks gas regime.

As a result, we can write the field operator as 
\begin{equation}
\hat{\Psi}(z,t)=\Phi (z,t)+\hat{\psi}(z,t),  \label{eq:field}
\end{equation}%
with $\Phi (z,t)$ the condensate wavefunction and $\hat{\psi}(z,t)$ the
quantum fluctuation. To the zero-th order of $\hat{\psi}(z,t)$, we obtain
the mean-field Gross-Pitaevskii (GP) equation
\begin{equation}
i\hbar \partial_t \Phi(z,t)= [H_0(t) +\bar{g} |\Phi(z,t)|^2] \Phi(z,t).
\label{eq:GP_supp}
\end{equation}
\\
\emph{Hartree-Fock-Bogoliubov approach.--- }Initially, our system is a true
TF condensate with negligible fluctuations. As the kick number increases,
the delocalization of the system may be accompanied with the rapid
proliferation of non-condensate atoms, especially for strong interactions. The
GP approach may apply only in the regime where the non-condensate particle number
is much smaller than the condensate atom number. Therefore, it is important to go beyond the mean-field approach and examine the excitation properties.
Here we employ the Hartree-Fock-Bogoliubov approach to consider the effects
of quantum fluctuation $\hat{\psi}(z,t)$~\cite{grif96}. Substituting Eq.~\ref{eq:field} into Eq.~\ref{eq:nlse}, and keeping the terms up to the second order
of $\hat{\psi}(z,t)$, we obtain 
\begin{equation}
i\hbar \partial _{t}\Phi (z,t)=[H_{0}(t)+\bar{g}|\Phi (z,t)|^{2}+2\bar{g}%
n^{\prime }]\Phi (z,t)+\bar{g}m^{\prime \ast }(z,t),
\end{equation}%
with $n^{\prime }=\langle \hat{\psi}^{\dag }(z,t)\hat{\psi}(z,t)\rangle $, $%
m^{\prime }=\langle \hat{\psi}(z,t)\hat{\psi}(z,t)\rangle $. The quantum
fluctuation satisfies 
\begin{equation}
i\hbar \partial _{t}\hat{\psi}(z,t)=[H_{0}(t)+2\bar{g}(|\Phi |^{2}+n^{\prime
})]\hat{\psi}(z,t)+\bar{g}(|\Phi |^{2}+m^{\prime })\hat{\psi}^{\dag }(z,t).
\end{equation}

Using the Bogoliubov transformation 
\begin{equation}
\hat{\psi}(z,t)=\sum_{j}u_{j}(z,t)\hat{\beta}_{j}-v_{j}^{\ast }(z,t)\hat{%
\beta}_{j}^{\dag },
\end{equation}%
with $\hat{\beta}_{j}^{\dag }$ the excitation creation operator, we obtain
the Bogoliubov equation 
\begin{equation}
i\hbar \partial _{t}\left( 
\begin{array}{cc}
u_{j} &  \\ 
v_{j} & 
\end{array}%
\right) =\left( 
\begin{array}{cc}
M_{0} & M_{1} \\ 
-M_{1}^{\ast } & -M_{0}%
\end{array}%
\right) \left( 
\begin{array}{cc}
u_{j} &  \\ 
v_{j} & 
\end{array}%
\right) ,
\end{equation}%
where $M_{0}=H_{0}(t)+2\bar{g}(|\Phi |^{2}+n^{\prime })$ and $M_{1}=\bar{g}%
(|\Phi |^{2}+m^{\prime })$. We are only interested in the stability of the
system (\textit{i.e.}, whether the non-condensate particle number $n^{\prime }$
increases exponentially), which can be determined even when $n^{\prime }$ is much
smaller than the condensate. Therefore, we can ignore the $n^{\prime }$ and $%
m^{\prime }$ in the above equations for simplicity~\cite{zhan04}.\\

\emph{Mapping to nonlinear Anderson model.---}As discussed above, in the
regime where the non-condensate atom number is small, the GP equation gives a good
description of the nonlinear QKR. It is known that the non-interacting QKR
can be mapped to the Anderson model, i.e, a one-dimensional lattice in
momentum space with on-site disorders~\cite{fish82}. Here we briefly review
the mapping and discuss the effects of interactions. The QKR is a Floquet
system with $\mathcal{H}(t)=\mathcal{H}(t+T)$, therefore the wave function
takes the form of $\Phi (z,t)=e^{-i\epsilon t}\phi (z,t)$ with periodic part 
$\phi (z,t)=\phi (z,t+T)$ and quasienergy $\epsilon$. Denoting $\phi _{\pm }(z)$ as the wave functions
just after and before the kick, we have 
\begin{equation}
\phi _{+}(z)=e^{i{\kappa }\cos {2k_{L}z}}\phi _{-}(z).
\end{equation}%
For $g=0$, the free evolution between kicks yields 
\begin{equation}
\phi _{-,j}=e^{i(\epsilon T-4j^{2}E_{\text{rec}}T/\hbar )}\phi _{+,j},
\end{equation}%
where $\phi _{\pm ,j}=\frac{1}{\sqrt{Z}}\int dze^{i2jk_{L}z}\phi _{\pm }(z)$
are the $j-$th Fourier components of the wave function with system
size $Z$. For simplicity we have neglected the harmonic trap. From the above two equations, we obtain the effective Anderson model 
\begin{equation}
V_{j}\bar{\phi}_{j}+\sum_{j^{\prime }\neq 0}K_{j^{\prime }}\bar{\phi}%
_{j+j^{\prime }}=\omega \bar{\phi}_{j},  \label{eq:Anderson}
\end{equation}%
with on-site disorder $V_{j}=\tan {(\epsilon T/2-2j^{2}E_{\text{rec}}T/\hbar
)}$, hopping rates $K_{j}=\frac{1}{\sqrt{Z}}\int dze^{i2jk_{L}z}\tan [\frac{{%
\kappa }}{2}\cos (2k_{L}z)]$, and energy $\omega =-K_{0}$. Here $\bar{\phi}%
_{j}=({\phi }_{-,j}+{\phi }_{+,j})/2$.

In the presence of interactions, the free evolution between two kicks can be
written as 
\begin{equation}
i\hbar \partial _{t}\phi (z,t)=[\frac{p^{2}}{2m}-\hbar \epsilon +\bar{g}%
|\phi |^{2}]\phi (z,t).
\end{equation}%
In the momentum space, 
\begin{equation}
i\hbar \partial _{t}\phi _{j}(t)=\sum_{j^{\prime }}\left[ \left( \frac{%
4j^{2}\hbar ^{2}k_{L}^{2}}{2m}-\hbar \epsilon \right) \delta _{j^{\prime
},0}+\frac{\bar{g}}{Z}\sum_{j_{1}}\phi _{j_{1}}^{\ast }\phi _{j_{1}-j^{\prime }}\right]
\phi _{j+j^{\prime }}(t).
\end{equation}%
We see that the contact interactions in real space become long-range
interactions in momentum space, which lead to the delocalization in
momentum space, as observed in experiments. We can split the interactions
into two parts: the diagonal interaction with $j^{\prime }=0$ or $j^{\prime
}=j_{1}-j$, and the off-diagonal interaction with $j^{\prime }\neq 0$ and $%
j^{\prime }\neq j_{1}-j$. The free evolution becomes 
\begin{eqnarray}
i\hbar \partial _{t}\phi _{j}(t) &=&\left( \frac{4j^{2}\hbar ^{2}k_{L}^{2}}{%
2m}-\hbar \epsilon \right) \phi _{j}(t)  \notag \\
&+&\frac{\bar{g}}{Z}\left( 2N_{\text{atom}}-|\phi _{j}|^{2}\right) \phi _{j}(t)  \notag
\\
&+&\frac{\bar{g}}{Z}\left[ \sum_{j^{\prime }\neq 0,j_{1}-j}\phi _{j_{1}}^{\ast }\phi
_{j_{1}-j^{\prime }}\right] \phi _{j+j^{\prime }}(t).
\end{eqnarray}%
The first line on the right-hand side of the above equation corresponds to
the single-particle evolution, while the second and third lines correspond
to diagonal interactions (which lead to on-site attraction) and off-diagonal
interactions (which lead to infinite long-range hopping). 

Due to the  infinite long-range
feature, it is hard to obtain the relation between $\phi _{-}$ and $\phi
_{+}$ based on the non-linear free evolution. However, if we consider
only the diagonal interaction, we can obtain the solution 
\begin{equation}
\phi _{-,j}=e^{i[\epsilon T-4j^{2}E_{\text{rec}}T/\hbar +\bar{g}(2N_{\text{%
atom}}-|\phi _{+,j}|^2)T/Z\hbar ]}\phi _{+,j}.
\end{equation}%
The nonlinear Anderson model takes the same form as Eq.~\ref{eq:Anderson},
while the on-site disorder becomes nonlinear with $V_{j}=\tan {(\epsilon
T/2-(2j^{2}E_{\text{rec}})T/\hbar -\bar{g}N_{\text{atom}}T/Z\hbar +\bar{g}%
T|\sum_{j^{\prime }}\bar{\phi}_{j+j^{\prime }}(K_{j^{\prime }}+\delta
_{j,j^{\prime }})|^{2}/2Z\hbar )}$. Though there is no analytical expression, the off-diagonal interaction induces
effective infinite long-range hopping in the nonlinear Anderson model, which together with the diagonal interaction are responsible for the dynamical
delocalization.

\section{Numerical Results}

For our numerical simulation, we take $2\hbar k_{L}$ and $T$ as the momentum and time units. The GP equation becomes 
\begin{equation}
   i\kbar \partial_\tau \Phi(\theta,\tau)= 
   [-\frac{\kbar^2\partial_\theta^2}{2} + {V}(\theta) - K \cos(\theta) \sum_{n_p} \delta(\tau-n_p) +g |\Phi(\theta,\tau)|^2] \Phi(\theta,\tau),
   \label{eq:GP_dl}
\end{equation}
where $\tau=t/T$, $\theta=2k_L z$, 
$\kbar=8TE_\text{rec}/\hbar$ and 
$K=\kbar{\kappa}=\kbar V_\text{kick}t_p/2\hbar$ are dimensionless parameters. The harmonic confinement $ V(\theta)=\frac{1}{2}{\omega}_\theta^2\theta^2$ is expressed in terms of the dimensionless trapping frequency ${\omega}_\theta=\omega_zT$.
The dimensionless interaction strength is expressed as ${g}=2\bar{g}k_LT\kbar/\hbar=\kbar^2\frac{k_La_s}{(k_La_\perp)^2}$, and the wave function is normalized as $\int d\theta |\Phi(\theta,\tau)|^2 =N_\text{atom}$ where $N_{\text{atom}}$ is the atom number per tube. 

 For $^{174}$Yb atoms, $k_{L}=2\pi /\lambda $ with $\lambda =1073$ nm, and 
we have $E_\text{rec}=\frac{\hbar^2k_L^2}{2m}=\hbar \times 2\pi\times 1$ kHz and $a_{s}=5.55$ nm. The effective Planck
constant $\kbar=5.26$ for $T=105 \mu$s. The kick
strength is $K\simeq 2.63$ (i.e., ${\kappa }\simeq 0.5$) for $V_{\text{kick}%
}=80E_{\text{rec}}$ and $t_{p}=2 \mu $s. If we consider the peak atom density
of the center tube to be about $\bar{n}_{\rm 1D}\simeq 25/\mu $m, as observed
experimentally, then the peak interaction strength would be $\bar{g}\bar{n}%
_{\rm 1D}\simeq 5.8E_{\text{rec}}$ for the transverse trap $a_{\perp }\simeq 53$%
 nm with $s_{\perp }=106$. For the dimensionless parameters, we have ${g}%
\simeq 9$ and peak density $n_{\rm 1D}=|\Phi (0,0)|^{2}\simeq \bar{n}%
_{\rm 1D}/2k_{L}\simeq 2.1$, leading to the dimensionless center tube peak
interaction strength ${g}n_{\rm 1D}\simeq 19$. If we consider a TF condensate in
the trap, this value corresponds to the atom number $N_{\text{atom}}\simeq
400$ for the center tube, with FWHM about 18$\mu $m. On the other hand, atom
numbers in the tubes decrease slightly with the distance away from the
center, and we consider only tubes within the half-maximum transverse circle
(atom numbers decay quickly beyond this half-maximum circle), then the
averaged atom number in each tube would be $N_{\text{atom}}\simeq 300$
(notice that the TF condensates in all tubes have the same chemical
potential). In our numerical simulation of the 1D dynamics, the initial atom number is fixed at $N_{\text{atom}}=300$ to match the observed center-tube
peak atom density in the experiment (i.e., $\sim 25$ atoms$/\mu $m for $%
s_{\perp }=106$). We use a Gaussian trap that
is the same as that in the experiment, whose low-energy small $\theta $
expansion leads to the harmonic trap $V(\theta )$.

We solve the GP equation numerically using the time-split method, with the initial state obtained by the imaginary time evolution of the GP equation without kicks. The results are shown in Fig.~3 of the main text. Here we adopt the experimental parameters given in Fig.~3a and b of the main text. 
Our numerical simulations show good agreement with the experimental data for parameters deep in the delocalization/localization regime, while around the localization-delocalization phase boundary, we observe large deviations between the simulation and experiment. This is because the
dynamics become more sensitive to the parameters and the initial states
(e.g., the numerical simulation starts with the ground state of the BEC
while the initial state in the experiment may contain a small portion of
excited states). Moreover, several strongly interacting center tubes may
enter the delocalized phase while the surrounding tubes are still in the
localized phase in experiments, therefore the numerical simulations using
the averaged interaction strength may not match the experimental data near
the phase boundary. 

The numerical phase diagram in the $K$-$gn_{\text{1D}}$ plane is shown in Figure \ref%
{figs4} (same as Fig. 3c of the main text). Here the phase transition
points marked by the blue and red squares are obtained by solving the GP and Hartree-Fock-Bogoliubov equations, respectively. The phase boundaries marked by the blue solid and red dashed lines are obtained by fitting to the phase transition points. Due to the dynamical localization in the absence of interaction, the phase transition point goes to infinity as $g$ decreases to zero. We find that the critical kick strength $K_{c}$ shows a roughly linear dependence on $%
(gn_{\text{1D}})^{-1}$, therefore we use the ansatz $y=ax^{-1}+b$ to fit the numerical results. 
The fitting parameters are $a=24.95$ and $b=-0.12$ for the blue solid line, and $%
a=26.34$ and $b=0.14$ for the red dashed line. Above 
the blue solid line, the mean energy starts to increase with the pulse number, as
shown in Figure \ref{figs5}, indicating the dynamical delocalization. The
red dashed line corresponds to the boundary between stable and unstable regimes. 
Above the red dashed line, the BEC becomes unstable as kick number increases, which
is manifest by the exponential increase of noncondensed atom number, as
shown in Figure \ref{figs6}. 
The critical kick strengths ($K_{c}$) for both phase transitions
decrease with $gn_{\text{1D}}$. 
We notice that the two phase boundaries are close to each other, which
suggests that the dynamical delocalization is accompanied by the instability of
the BEC. 


As discussed above, only $gn_{\text{1D}}$ matters in the dynamics. We take
the peak density of the center tube to be $25/\mu $m for $s_{\perp }=106$,
consistent with the experiment. We also take into account the averaging
effects over different tubes, which lead to an effective atom number $N_{%
\text{atom}}=300$ in one tube. For a given interaction $g$, the density $n_{%
\text{1D}}$ can be obtained by assuming a TF gas, which is consistent with
the ground state obtained using imaginary-time GP equation. Larger experimental values for average atom number and system axial size imply that we do not start from the exact ground state. This is consistent with the small departures from adiabaticity in loading the BEC into the two-dimensional optical lattice and the finite non-condensate fraction, as discussed earlier. In the numerical simulation of the experimental data in Fig.~3a,b and Fig.~4 of the main text, we also take into account the finite trap depth which leads to atom loss for $\langle E_{z}\rangle $ above $10E_{\mathrm{rec}}$. In the numerical simulation for the 3D case, $gn_{\rm 1D}$ is obtained by adjusting $a_{\perp }$ in Eq.~\ref{eq:GP_dl} to match the measured chemical potential.





\begin{figure}[ht]
		\center
		\includegraphics[width=0.5\textwidth]{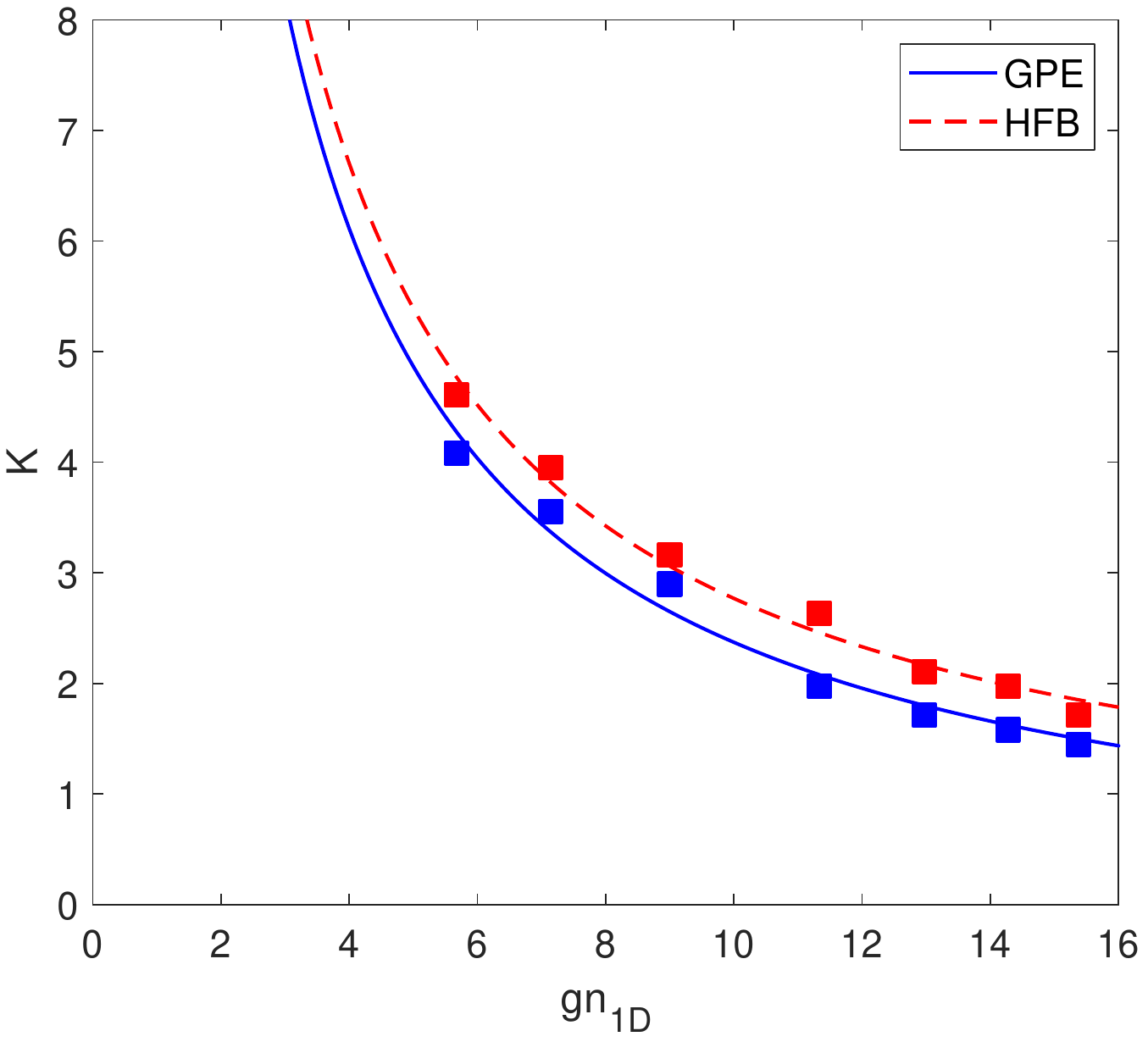}
\caption{{\bf Phase diagram of the localization behavior of the system in the space of kick strength and interaction strength.} The blue and red markers are the phase transition points obtained by solving the GP and Hartree-Fock-Bogoliubov equations, respectively. The system is localized and stable below the markers, and delocalized and unstable above the markers.
The phase boundaries marked by the blue solid and red dashed lines are the same as those in Fig. 3c of the main text and are obtained by fitting to the markers. Here we use the ansatz $y=ax^{-1}+b$ for the fitting. The fitting parameters are $%
a=24.95$ and $b=-0.12$ for the blue solid line, and $a=26.34$ and $b=0.14$ for the
red dashed line.}
\label{figs4}
\end{figure}

\begin{figure}[ht]
\center
\includegraphics[width=0.76\textwidth]{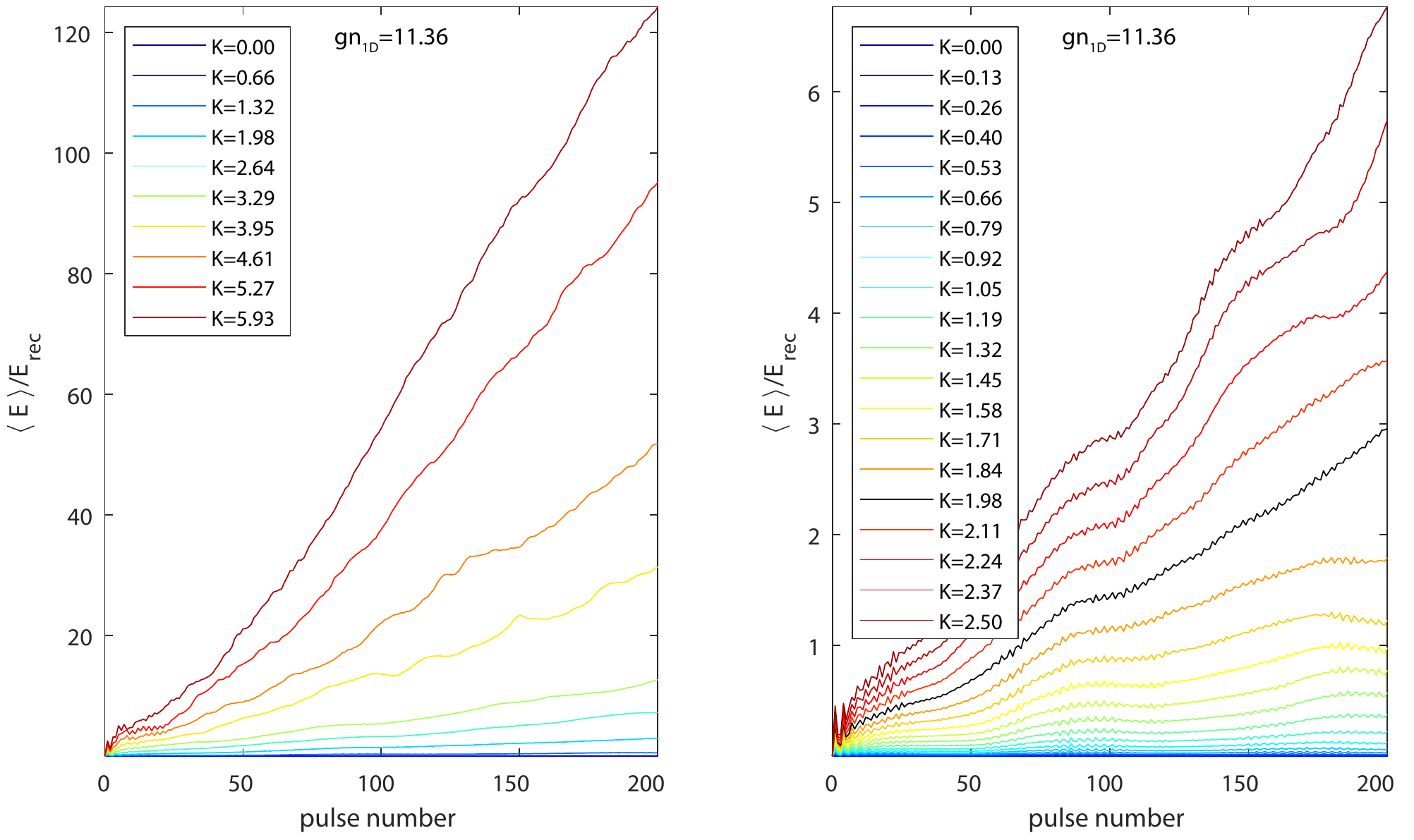}
\caption{{\bf Mean energy as a function of pulse number for different $K$ with $gn_\text{1D}=11.36$.} In general, the energy increases faster for stronger kicks $K$ (left
panel). In the localized phase, the energy saturates at a small
value as the kick number increases, while in the delocalized phase, the
energy increases diffusively with the kick number (right panel). Here 
$\kbar=5.26$. The black line marks the critical kick strength $K_c$%
.}
\label{figs5}
\end{figure}

\begin{figure}[ht]
\center
\includegraphics[width=0.76\textwidth]{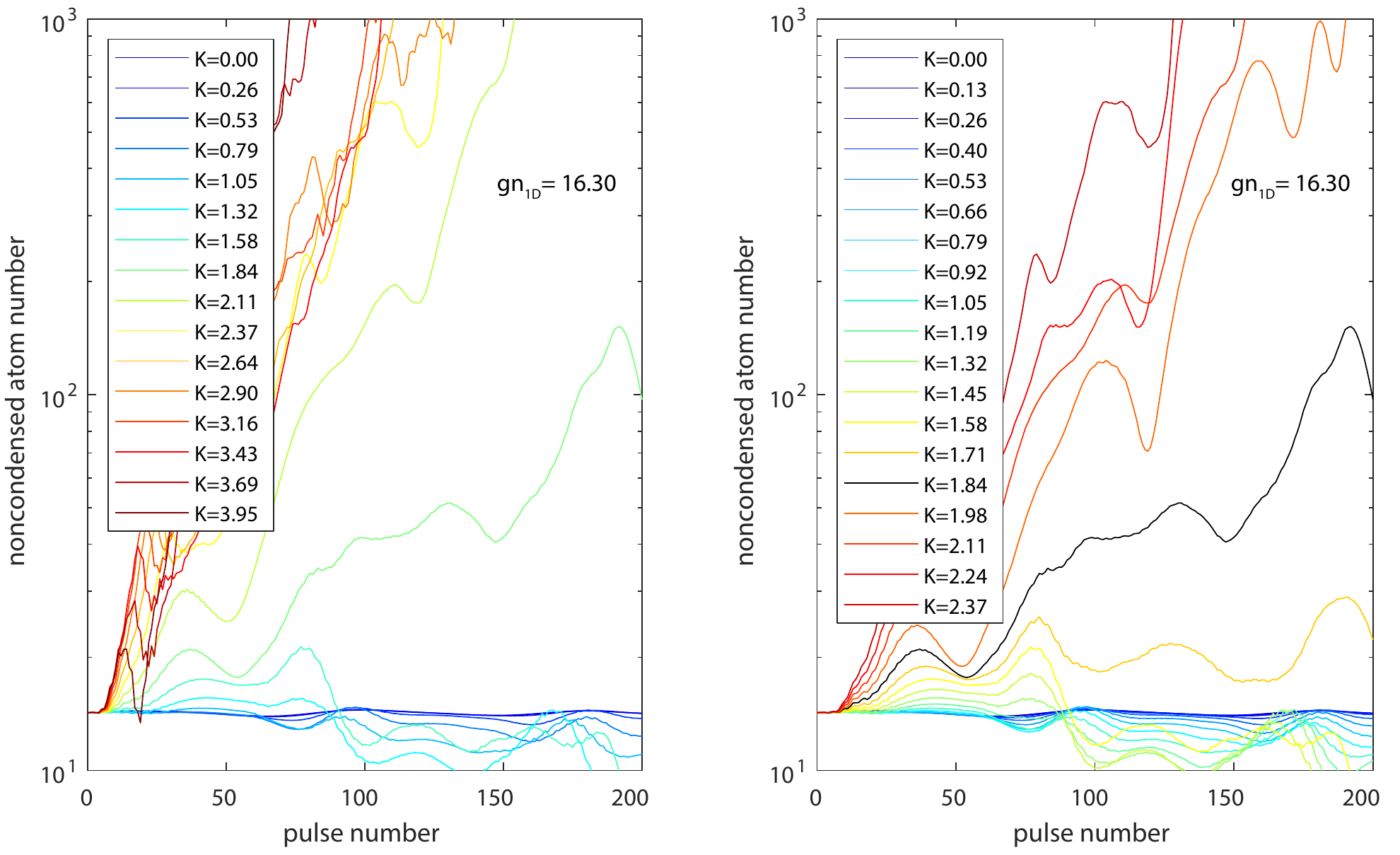}
\caption{{\bf Noncondensed atom number as a function of pulse number for different $K$ with $gn_{\rm 1D}=16.3$.} At large kick number, the BEC becomes unstable as the interaction strength increases. The transition to instability happens around the transition to delocalization. Here the black line marks the critical kick strength $K_{c}$.}
\label{figs6}
\end{figure}


\end{document}